\documentclass[twocolumn,preprintnumbers,amsmath,amssymb,showpacs]{revtex4}

\usepackage{graphicx}% Include figure files
\usepackage{dcolumn}% Align table columns on decimal point
\usepackage{bm}% bold math
\usepackage[final]{pdfpages}

\begin{document}
\title{S-wave Superconductivity in Optimally Doped SrTi$_{1-x}$Nb$_x$O$_3$ Unveiled by Electron Irradiation }
\author{Xiao Lin,$^{1}$ \footnote{xiao.lin@espci.fr}  Carl Willem Rischau,$^{1,2}$ Cornelis J. van der Beek,$^{2}$ Beno\^{\i}t Fauqu\'e,$^{1}$ and Kamran Behnia$^{1}$}
\affiliation{(1) Laboratoire de Physique et d'Etude des Mat\'{e}riaux (CNRS/ESPCI/UPMC), Paris, F-75005, France\\
(2) Laboratoire des Solides Irradi\'{e}s (CNRS-CEA/DSM/IRAMIS), Ecole Polytechnique, 91128 Palaiseau cedex, France \\}

\date{\today}

\begin{abstract}
We report on  a study of electric resistivity and magnetic susceptibility measurements in electron irradiated SrTi$_{0.987}$Nb$_{0.013}$O$_3$ single crystals.  Point-like defects, induced by electron irradiation, lead to an almost threefold enhancement of the residual resistivity, but barely affect the superconducting critical temperature ($T_c$). The pertinence of Anderson's theorem provides strong evidence for a s-wave superconducting order parameter. Stronger scattering leads to a reduction of the effective coherence length ($\xi$) and the deduced coherence length in the clean limit ($\xi_0$) is around the BCS coherence length ($\xi_{BCS}$). Combined with thermal conductivity data pointing to multiple nodeless gaps, the current results identify optimally doped SrTi$_{1-x}$Nb$_x$O$_3$ as a multi-band s-wave superconductor.
\end{abstract}


\maketitle

\section{\label{sec:level1}Introduction}

Scattering mixes the superconducting order parameter at separate points on the Fermi surface. As a consequence, one can probe changes in the two-particle wave-function by tuning disorder. Its effect on the superconducting transition provides an opportunity to explore the symmetry of the superconducting gap. According to  Anderson's theorem, in a conventional s-wave superconductor the critical temperature ($T_c$) is insensitive to nonmagnetic disorder \cite{Anderson}. On the other hand, in superconductors with non-trivial gap symmetry, e.g., cuprates \cite{Uchida,Tolpygo,Albenque}, Sr$_2$RuO$_4$ \cite{Mackenzie}, and heavy fermions \cite{Stewart}, $T_c$ is extremely sensitive to potential scattering and the superconducting ground state can be completely destroyed by disorder \cite{Radtke,Abrikosov1,Varma,Borkowski}. In multi-band superconductors such as MgB$_2$ and iron pnictides, interband scattering rather than intraband scattering plays a key role in suppressing $T_c$ and the effect of disorder depends on the ratio of interband to intraband scattering matrix elements \cite{Prozorov,Golubov,WangY}.

Chemical substitution can be used to introduce disorder. In cuprates, $T_c$ is drastically suppressed by Zn doping, providing strong evidence for d-wave symmetry \cite{Uchida}. Particle irradiation provides an alternative avenue of creating artificial defects without introducing any foreign ions. In YBa$_2$Cu$_3$O$_{7-\delta}$, scattering induced by electron irradiation suppressed $T_c$ in a manner similar to Zn substitution \cite{Albenque,Albenque1,Legris}. On the other hand, in the s-wave superconductor MgB$_2$, superconductivity is robust with respect to electron irradiation \cite{Blinkin,Klein,Blinkin1}. In present paper, electron irradiation is utilized to investigate the superconducting order parameter in optimally doped  SrTi$_{1-x}$Nb$_x$O$_3$ single-crystals.

%However, neutron and $\alpha$-particle irradiation of MgB$_2$ led to an apparent suppression of $T_c$ \cite{Wilke,Gandikota,Putti1}. The shape and size of defects, which influence scattering, depend on the type of irradiation.Energetic heavy ions generate columnar defects along the ion trajectories \cite{Civale,Gerhauser,Nakajima1}. Protons, $\alpha$-particles, and neutrons most likely produce defect clusters of nm size \cite{Prozorov}. On the other hand, high energy electrons (1$-$10 MeV) generate point-like defects in the form of interstitial-vacancy pairs (Frenkel pairs) \cite{Damask}. This makes electron irradiation a suitable method for introducing controlled disorder.

A band insulator with an energy gap of 3.2 eV, SrTiO$_3$ is close to a ferroelectric instability aborted due to quantum fluctuations \cite{Muller}. Its huge permittivity at low temperature leads to a very long Bohr radius and a precocious metallicity. Three conducting bands, originating from Ti t$_{2g}$ orbits and centered at the $\Gamma$ point can be successfully filled by n-doping \cite{Marel}. A superconducting dome, with a peak  $T_c \simeq$ 450 mK \cite{Schooley,Schooley1,Schooley2,Xiao,Xiao1} exists between charge carrier densities of 3$\times$10$^{17}$ to 3$\times$10$^{20}$ cm$^{-3}$. In case of an extremely low carrier concentration, superconductivity still survives with the Fermi energy ($\epsilon_F$) lower than the Debye temperature ($T_D$) \cite{Xiao}, which challenges the conventional phonon mediated weak coupling BCS theory. Some exotic superconducting mechanisms have been proposed to explain the superconductivity in SrTiO$_3$ \cite{Takada, Appel, Rowley}.

The symmetry of the superconducting order parameter has been barely explored in this system. In 1980, Binnig and co-authors detected two distinct superconducting gaps by planar tunneling measurements \cite{Binnig}. However, a recent tunneling experiment on the superconducting LaAlO$_3$/SrTiO$_3$ interface did not detect multiple gaps \cite{Mannhart}. More recently, thermal conductivity measurements found multiple nodeless gaps in optimally doped SrTi$_{1-x}$Nb$_x$O$_3$ single crystals, paving the way for the identification of the symmetry of the superconducting order parameter \cite{Xiao2}.  A latest study reported the existence of electron pairs well beyond the superconducting ground state in quantum dots fabricated on the LaAlO$_3$/SrTiO$_3$ interface \cite{Levy}. In this paper, we present a study of ac susceptibility and resistivity in SrTi$_{1-x}$Nb$_x$O$_3$ irradiated with high energy electrons and provide unambiguous evidence for s-wave superconductivity.

%We also find an intriguing signature of  nonlocal electrodynamics, which may be related to the long effective Bohr radius of the parent insulator of this dilute superconductor.

\section{\label{sec:level1}Experimental}

The SrTi$_{1-x}$Nb$_x$O$_3$ (x=0.013) single crystals used in this study were obtained commercially as the one used in thermal conductivity measurements \cite{Xiao2}. Four samples with size of 5$\times$2.5$\times$0.5 mm have been cut from the same single crystal and gold was evaporated on their surface to make Ohmic contacts. Three of them were irradiated with 2.5 MeV electrons at the SIRIUS accelerator facility of the Laboratoire des Solides Irradi\'{e}s. Irradiations were performed at 20 K in liquid hydrogen to obtain a uniform distribution of point defects in the material. After irradiation, the samples were stored in liquid nitrogen to avoid room temperature annealing of the irradiation-induced defects. The resistivity and Hall effect around the superconducting transition temperature were measured with a standard four probe method in a dilution refrigerator within a few days after the irradiation. The transport properties were rechecked in a Quantum Design PPMS system above 2 K a few months later. The Hall carrier density and residual resistivity have barely changed with time. Gold contacts that are large compared to the size of the samples may give rise to an uncertainty of 10$\%$ in the transport measurements. Finally, the ac susceptibility was measured in a homemade set-up, which consisted of one primary field coil and one compensating pick-up coil with two sub-coils with their turns in opposite direction. The exciting ac current was supplied and the induced voltage signal was picked up by a Lock-in amplifier. The applied ac magnetic field was as low as 10 mG, with frequencies between 2000 and 4000 Hz.

\section{\label{sec:level1}Results and Discussion}

\begin{figure}\resizebox{!}{0.5\textwidth}
{\includegraphics{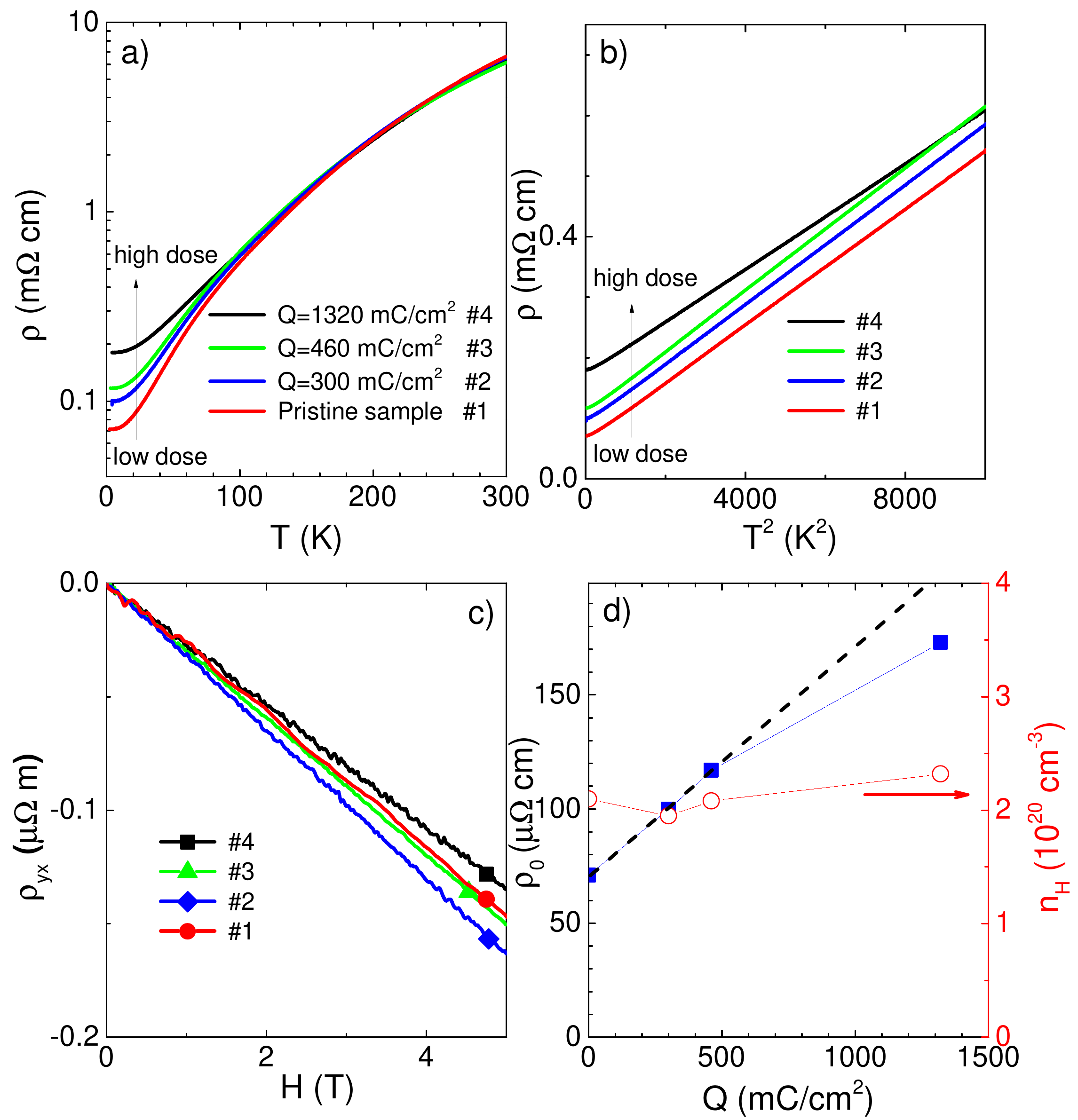}}
\caption{Resistivity and Hall coefficient in pristine and electron-irradiated SrTi$_{0.987}$Nb$_{0.013}$O$_3$ single crystals. a) Temperature dependence of resistivity (note the vertical log scale). The low temperature resistivity monotonically increases with irradiation dose. b) Resistivity as a function of T$^2$. All the samples show T$^2$ resistivity hardly altered by electron irradiation.  c) Hall resistivity ($\rho_{yx}$) as a function of magnetic field at 10K. d) Residual resistivity ($\rho_0$=$\rho(2K)$) and Hall carrier concentration ($n_H$) as a function of irradiation dose ($Q$). Irradiation enhances the residual resistivity by a factor of 2.5, but leaves the carrier density virtually unchanged ($n_H$ $\approx$ 2.1$\times$10$^{20}$ cm$^{-3}$). The dashed line is a guide to the eyes.}
\end{figure}

Fig. 1(a) and (b)shows the temperature dependence of the resistivity of the pristine sample No.1 and of samples No.2, No.3 and No.4 that were irradiated to total electron doses $Q$ = 300, 460 and 1320 mC/cm$^2$ respectively. Rather than modifying the room temperature resistivity, the electron irradiation induced defect scattering clearly increases the low temperature resistivity.
The residual resistivity $\rho_{0}=\rho(2K)$ amounts to 71 $\mu\Omega$.cm in the pristine sample and increases with increasing irradiation dose.
Consistent with ref. \cite{Xiao4}, all the samples present Fermi liquid behavior with T$^2$ resistivity expressed by $\rho=\rho_0+AT^2$. The T$^2$ prefactor $A$ from inelastic electron scattering is around 0.048 $\mu\Omega$.cm/K$^2$ with an error of 10$\%$. Hence the point-like defects induced by the electron irradiation barely affect the inelastic scattering at higher temperature, but only increase the elastic scattering at zero temperature.

Fig. 1(c) plots the Hall resistivity as a function of the magnetic field at 10 K. The Hall carrier concentration ($n_H$) plotted in Fig. 1(d) remains around 2.1$\times$10$^{20}$ cm$^{-3}$ with an error of 10$\%$, deduced from $R_H=1/n_He$ where $R_H=\rho_{yx}/B$ is the Hall coefficient. As seen in the figure, while the carrier concentration does not show any substantial change, $\rho_{0}$ increases linearly with the irradiation dose, indicating that the magnitude of the scattering rate is affected by the increased quantity of irradiation-induced scattering centers. $\rho_{0}$ amounts to 175 $\mu\Omega$.cm in sample No.4, enhanced by 104 $\mu\Omega$.cm compared to No.1, a magnitude comparable to what has been attained in other studies of impurity effects in superconductors such as cuprates \cite{Uchida} and pnictides \cite{Prozorov}. The mean-free-path ($l$) can be extracted using $l=\hbar\mu k_F/e$, where $\hbar$ and $e$ are the fundamental constants, $\mu$ is the Hall mobility and $k_F$ the Fermi wave factor, calculated from the carrier density assuming an isotropic single-component Fermi surface. With increasing $Q$, $l$ decreases from 50 to 19 nm.

\begin{figure}
\resizebox{!}{0.55\textwidth}
{\includegraphics{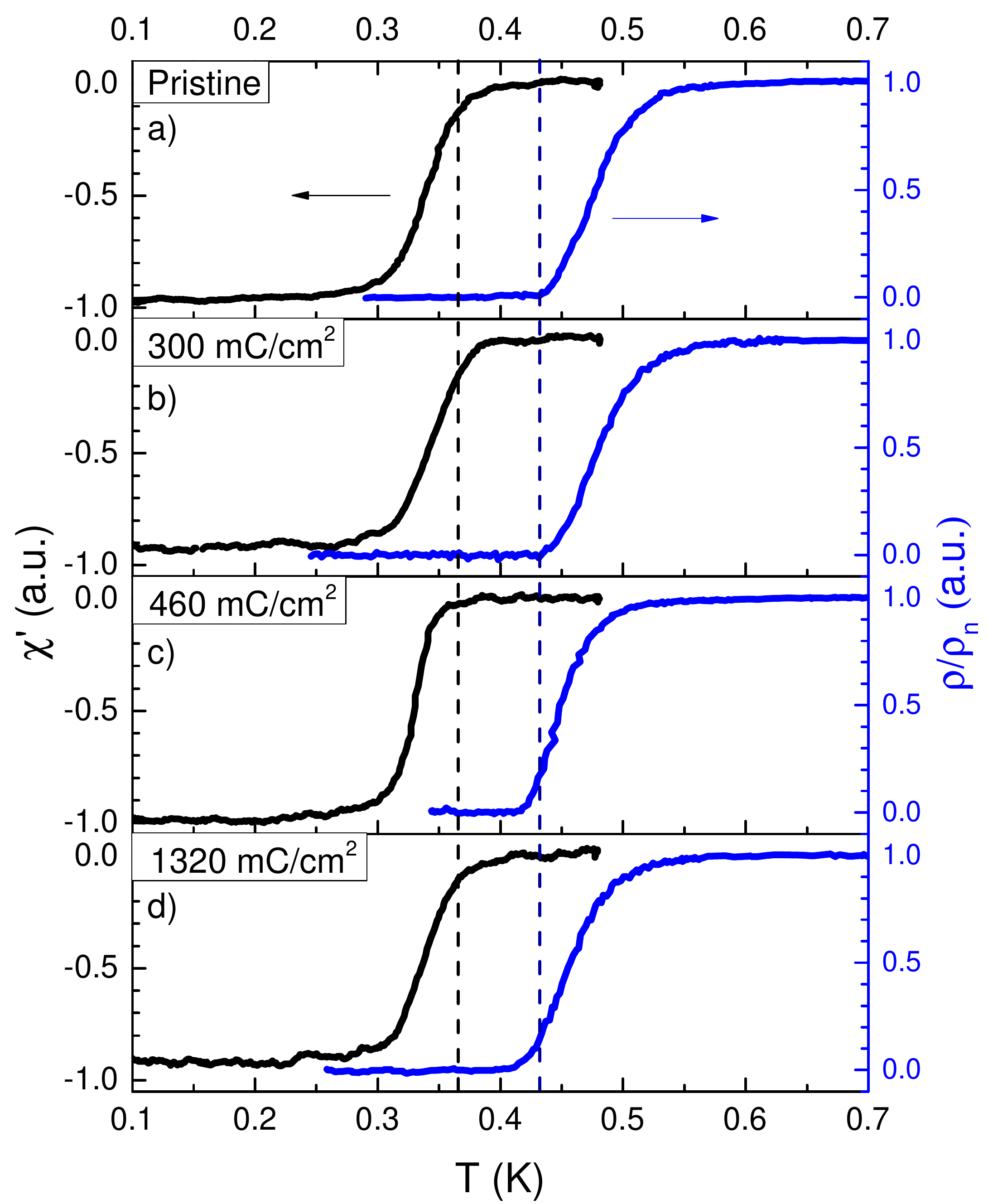}}
\caption{The real part of ac susceptibility ($\chi'$) and normalized resistivity ($\rho/\rho_n$) as a function of temperature around $T_c$ in absence of magnetic field for pristine and electron irradiated SrTi$_{0.987}$Nb$_{0.013}$O$_3$. Two vertical lines mark the transition temperatures in $\chi'$ and $\rho/\rho_n$. The superconducting transition barely shifts.}
\end{figure}

Fig. 2 shows  the superconducting transition in different samples such as observed through the real part of the susceptibility ($\chi'$) and the resistivity (normalized by its normal-state magnitude). There is a smooth transition in $\rho/\rho_n$ and the resistivity vanishes at a critical temperature ($T_{c-\rho}$) of 435 mK. On the other hand, $\chi'$ monitors bulk superconductivity, i.e., full flux exclusion. The bulk superconducting transition occurs at a temperature  $T_{c-\chi'}$, determined as the crossing point of two linear extrapolations, close to 370 mK. Such a difference of 65 mK between $T_{c-\rho}$ and $T_{c-\chi'}$, comparable to what was reported in our previous study comparing the specific heat, the thermal conductivity and the resistive superconducting transitions \cite{Xiao2}, is not changed by point-defect disorder. As seen in the figure, both $T_{c-\rho}$ and $T_{c-\chi'}$ remain basically the same in the four samples. This is the principal result of this study. In spite of the significant decrease of the charge-carrier mean-free-path, the critical temperature remains the same. Neither the width of the transition nor the superconducting shielding fraction are affected by the irradiations. Table 1 lists $T_{c-\rho}$ and $T_{c-\chi'}$.

\begin{figure}
\resizebox{!}{0.4\textwidth}
{\includegraphics{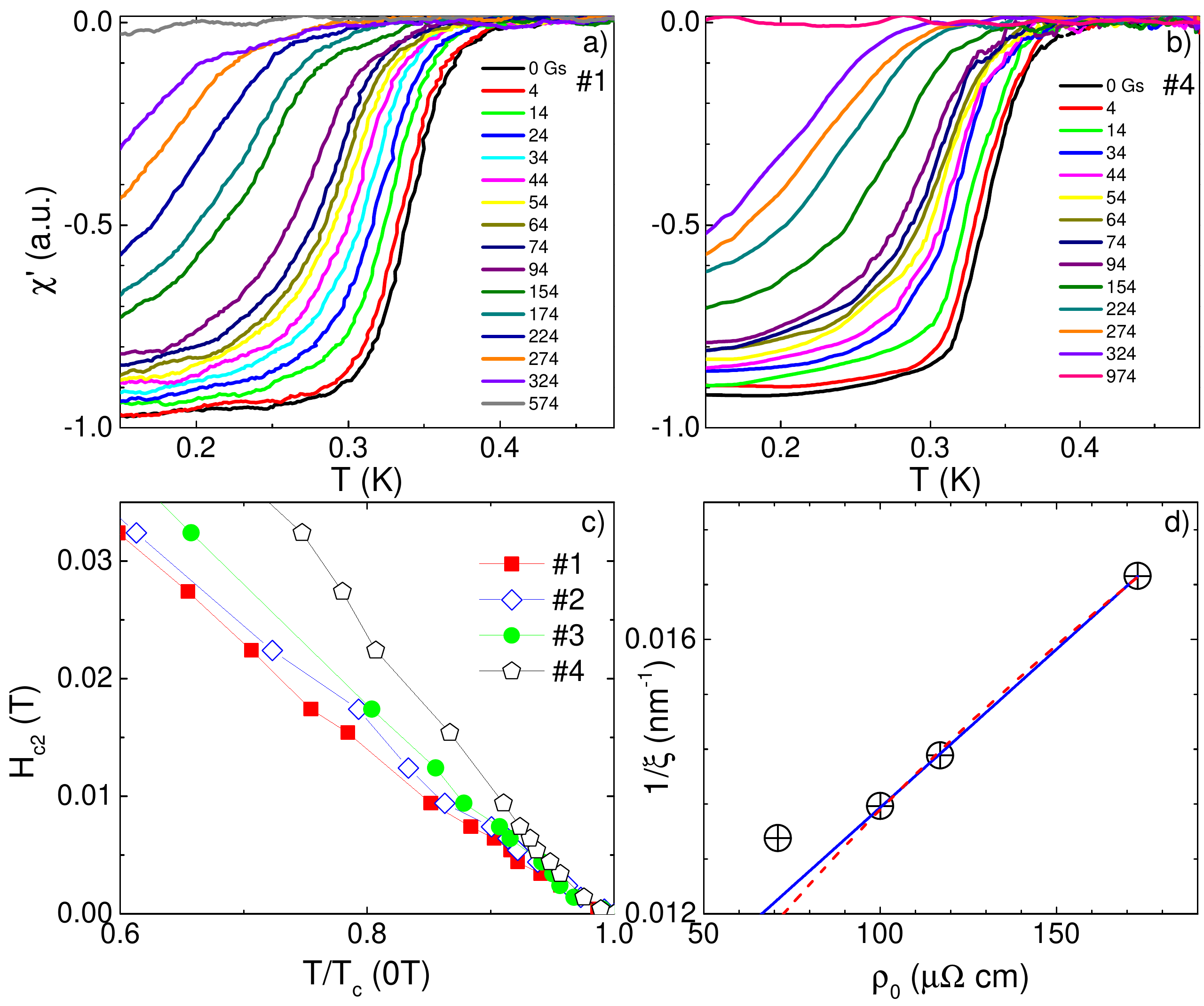}}
\caption{The evolution of the upper critical field ($H_{c2}$) and the effective coherence length ($\xi$) with electron irradiation. a) and b) $\chi'$ as a function of temperature around the superconducting transition at different magnetic fields, for samples No.1 and No.4 respectively. c) The evolution of $H_{c2}$ with $T_c/T_c(0T)$ from $\chi'$. The slope of $H_{c2}$ near $T_{c}$ evolves with irradiation. d) $1/\xi$, as extracted from upper critical field, as a function of $\rho_0$. The solid line is a linear fit from Eq. 2 and the dashed line is a fit from Eq. 3.}
\end{figure}

Figs. 3(a) and (b) plot $\chi'(T)$ near $T_c$ in presence of magnetic field for samples No.1 and No.4. As expected, the application of a magnetic field shifts the superconducting transition to lower temperatures. In Fig. 3(c), $H_{c2}$ is plotted as a function of $T/T_c(0T)$ for all the samples. A remarkable effect of the irradiation is to induce an enhancement of the slope of the upper critical field near $T_c$. One can quantify this effect by extracting the effective coherence length ($\xi$) from this slope using the expression based on the Werthammer-Helfand-Hohenberg theory \cite{WHH}:

\begin {equation}
1/\xi= \sqrt{\frac{2\pi\alpha}{\phi_0}T_c(0T)\frac{dH_{c2}}{dT}\mid_{T=T_c(0T)}}
\end {equation}

Here, $\phi_0$ is the flux quanta and $\alpha$ is a dimensionless parameter ranging from 0.725 in the clean limit to 0.69 in the dirty limit. By assuming a dirty superconductor, the effective coherence length passes from 76 nm in the pristine sample No.1 to 59 nm in the most irradiated sample No.4 (see Table 1). Shortening the mean-free-path leads to a decreasing effective coherence length $\xi$. This is expected, since $\xi$ can be expressed in BCS theory as:

\begin {equation}
1/\xi=1/\xi_0+1/\beta
\end {equation}

Here, $\xi_0$ is the intrinsic superconducting coherence length and $\beta$ is the characteristic length of electrodynamic response of the normal state current. Pippard argued that the order of magnitude of $\beta$ in a metal is the mean-free-path of electrons ($l$) \cite{Pippard,Tinkham}. Plotting  $1/\xi$ as a function of $\rho_0$ in Fig. 3(d), one can extract an intercept, which yields $\xi_0 \sim$ 112 nm. According to BCS and Ginzburg-Landau (GL) theory \cite{Tinkham}, Fig. 3(d) can alternatively be fitted by

\begin {equation}
1/\xi=\frac{2\sqrt{3}}{\pi}\frac{\sqrt{1+\xi_0/l}}{\xi_0}
\end {equation}
yielding $\xi_0 \sim$ 168 nm. Eq. 2 and 3 are valid only in the local condition, i.e. the vector potential (A) as well as the wave function ($\psi$) in BCS and GL theory varies slowly in a range of $\xi$, which requires $\xi$ much smaller than the penetration length ($\lambda$) \cite{Tinkham}. The local condition is satisfied in SrTi$_{1-x}$Nb$_x$O$_3$ in which H$_{c1}$ is two orders of magnitude smaller than H$_{c2}$ \cite{Ambler} and is strengthened by defect scattering induced by electron irradiation.
From both fits, $\xi_0$ is close to the BCS coherence length ($\xi_{BCS}$), which can be estimated to be $\xi_{BCS}=\hbar v_F/\pi\Delta(0) \sim$ 140 nm. The magnitude of the Fermi velocity, $v_F$ is given by  $\hbar k_F/m^*$ with $m^*=4m_e$ \cite{Xiao2}, while the superconducting gap $\Delta(0K) \sim80 \mu$eV is inferred from early tunneling experiments \cite{Binnig}. We conclude that $\xi_0$ is larger than the mean-free-path in all samples, indicating that the single crystals in this study are dirty superconductors \cite{xi}.

%Interestingly, the magnitude of $\beta$ derived using Eq. 2 is five to six times larger than the mean-free-path (see Table 1). This feature may be a peculiarity of this superconductor compared to those materials in which superconductivity emerges from a high carrier density metal. The huge electric permittivity in insulating SrTiO$_{3}$ leads to an effective Bohr radius (a$_B^*$), as long as 700 nm \cite{Xiao}, which is much larger than the mean-free-path. This may be the ultimate reason for a larger characteristic length for electrodynamic response in this low carrier density superconductor.

\begin{table}
 \renewcommand{\thetable}{1}
 \caption{Irradiation dose ($Q$), superconducting critical temperature from ac susceptibity ($T_{c-\chi'}$) and resistivity ($T_{c-\rho}$) at zero field, residual resistivity at 2K ($\rho_0$), T$^2$ prefactor ($A$) superconducting effective coherence length ($\xi$), and mean-free-path ($l$) for pristine and electron-irradiated SrTi$_{0.987}$Nb$_{0.013}$O$_{3}$ single crystals.}

 \begin{ruledtabular}
\begin{tabular}{lclccccl}

                        & No.1         & No.2           &  No.3     &No.4     \\

\hline

    $Q (mC/cm^2)$           & 0               & 300            &  460     & 1320     \\
% \hline
    % $n_H (10^{20} cm^{-3})$     & 2.1               & 1.95            &  2.08     & 2.32     \\
%\hline
    $T_{c-sus} (K) $           & 0.37            & 0.372          &  0.35    & 0.368     \\
    $T_{c-\rho} (K)$           & 0.435           & 0.435          &  0.42    & 0.419     \\
%\hline
    $\rho_0 (\mu\Omega.cm) $   & 71              & 100         & 117    & 173    \\
    $A (\mu\Omega.cm/K^2)$     &0.048            &0.049        &0.051  &0.043\\
%\hline

    $\xi (nm) $                & 76           &  74        & 70    & 59    \\
    $l (nm) $                  & 51           &  38        & 31    & 19    \\

\end{tabular}
\end{ruledtabular}
\label{TabI}
\end{table}

Let us compare our results with what has been reported in the case of other superconductors. Abrikosov and Gor'kov formulated a theory for the response of conventional superconductors  to magnetic impurities \cite{Abrikosov}. According to this theory, $T_c$ is suppressed, following:

\begin {equation}
-ln(\frac{T_c}{T_{c0}}) = \psi(\frac{1}{2}+\frac{\alpha T_{c0}}{4\pi T_c})-\psi(\frac{1}{2})
\end {equation}

Here, $\psi$ is the digamma function, $T_{c0}$ is the superconducting critical temperature in the clean limit, $\alpha=2\hbar\tau_s/k_BT_{c0}$ is the dimensionless pair breaking parameter and $\tau_s$ is the spin-flip scattering lifetime. Eq. 4 can be generalized to unconventional superconductors and their $T_c$ evolution with non-magnetic potential scattering. This can be done by replacing $\alpha$ with $\hbar\tau_p/k_BT_{c0}$, in which $\tau_p$ is the potential scattering lifetime \cite{Abrikosov1,Radtke,Tolpygo}. In order to make a simple comparison between experiment and theory, we take the residual resistivity as a measure of $\tau_p$, taken to be equal to the transport life time $\tau_{imp}$, expressed by $\tau_{imp}=\frac{m^*}{\rho ne^2}$.

\begin{figure}
\resizebox{!}{0.37\textwidth}
{\includegraphics{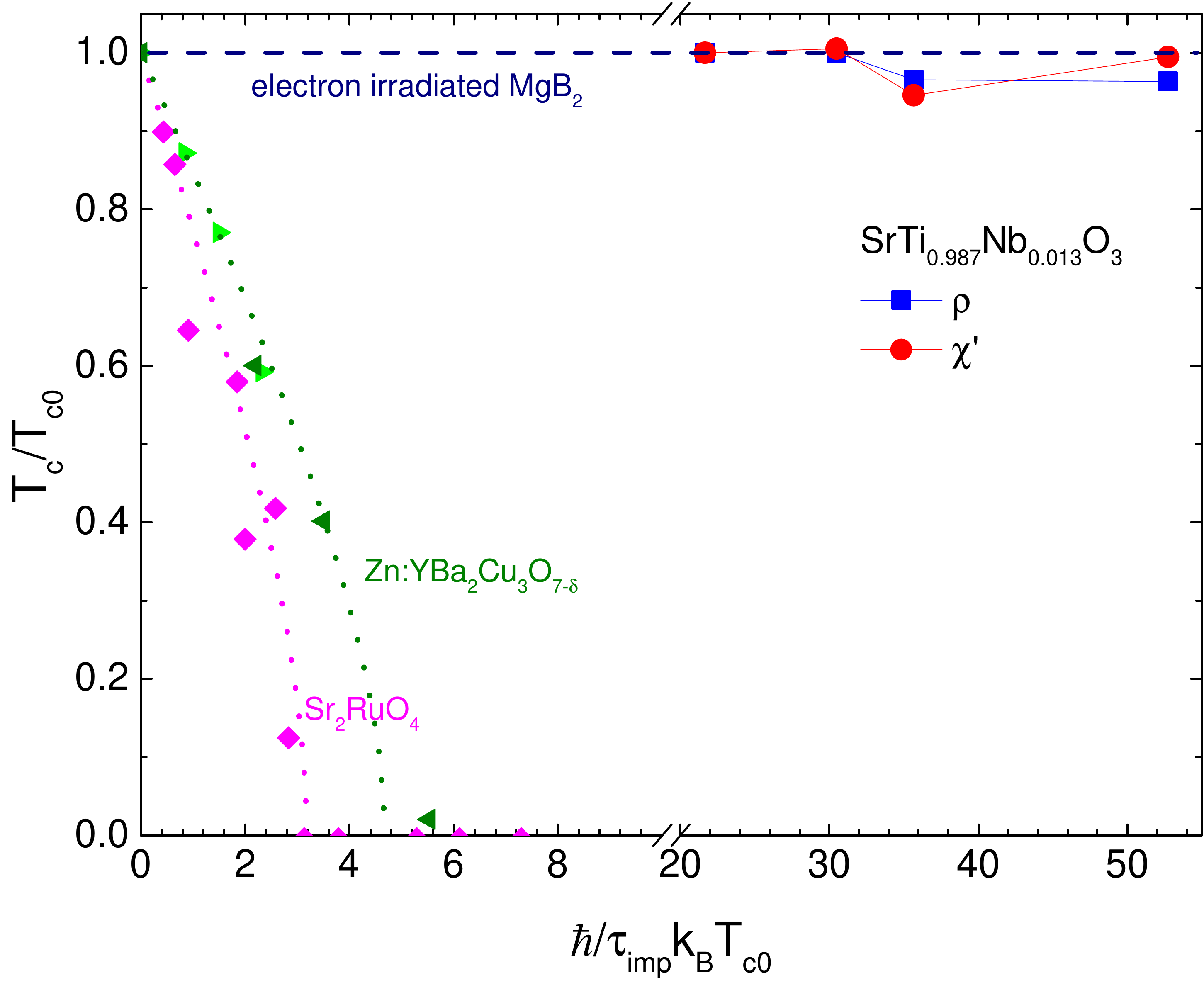}}
\caption{$T_c$/T$_{c0}$ as a function of the dimensionless pair-breaking rate $\alpha=\hbar\tau_{imp}/k_BT_{c0}$ in SrTi$_{0.987}$Nb$_{0.013}$O$_3$ determined from resistivity ($\blacksquare$) and ac susceptibility ($\bullet$). The data for MgB$_2$ under electron irradiation (the horizontal dashed line) is plotted for comparison, as well as those for two unconventional superconductors, Zn-doped cuprates ($\blacktriangleleft:$ YBa$_2$Cu$_3$O$_{6.63}$, $\blacktriangleright:$ YBa$_2$Cu$_3$O$_{6.93}$ ) and slightly disordered Sr$_2$RuO$_4$ ($\blacklozenge$). The dotted lines are guides to the eyes. Superconductivity is robust against impurity scattering in SrTi$_{0.987}$Nb$_{0.013}$O$_3$ and in MgB$_2$, but is rapidly suppressed in the two unconventional superconductors.}
\end{figure}

Fig. 4 shows $T_c/T_{c0}$ as a function of $\hbar\tau_{imp}/k_BT_{c0}$ ($\alpha$) for SrTi$_{0.987}$Nb$_{0.013}$O$_3$, compared with three other superconductors. These are the conventional superconductor MgB$_2$ \cite{Blinkin},  as well as two unconventional superconductors YBa$_2$Cu$_3$O$_{7-\delta}$ (d-wave) \cite{Uchida} and Sr$_2$RuO$_4$ (p-wave) \cite{Mackenzie}, which are both perovskites like the system under study. In both YBa$_2$Cu$_3$O$_{7-\delta}$ and Sr$_2$RuO$_4$, $T_c$ is extremely sensitive to the introduction of disorder and superconductivity is completely destroyed when $\alpha$ exceeds a number of the order of unity. In contrast, superconductivity in SrTi$_{0.987}$Nb$_{0.013}$O$_3$ is robust and $T_c$ shows a negligible variation even when $\alpha$ becomes very large. A similar behavior was observed in MgB$_2$. This is strong evidence for s-wave superconductivity in SrTi$_{0.987}$Nb$_{0.013}$O$_3$ and the main conclusion of this study.

\section{\label{sec:level1}Conclusion}

In summary, performing resistivity and ac susceptibility measurements on electron irradiated optimally doped SrTi$_{0.987}$Nb$_{0.013}$O$_3$, we have found that superconductivity  is robust against impurity potential scattering deep into the dirty limit ($\xi_0/l \backsim 5.9-8.8$). In addition, we have quantified the intrinsic clean coherence length ($\xi_0$) and found that it is comparable to the BCS coherence length ($\xi_{BCS}$). Combined with the thermal conductivity data, which pointed to the absence of nodal quasi-particles \cite{Xiao2}, this result identifies SrTi$_{1-x}$Nb$_x$O$_3$ as a multi-gap s-wave superconductor. The negligible suppression of $T_c$ also indicates that the relative weight of inter-band and intra-band scattering is not altered by electron irradiation.  In oxygen deficient SrTiO$_3$ with a carrier concentration 400 times lower than the samples studied here, the Fermi energy becomes one order of magnitude lower than the Debye temperature, a serious challenge for a phonon-mediated pairing mechanism \cite{Xiao}. Further experiments are required to probe the evolution of the gap symmetry and the pairing mechanism in a system whose superconductivity survives over three-orders-of-magnitude of carrier concentration.

\begin{acknowledgments}

We thank the staff of the SIRIUS accelerator for technical support. This work is supported by Agence Nationale de la Recherche as part of SUPERFIELD and QUANTUM LIMIT projects.
\end{acknowledgments}


\begin{thebibliography}{}
\bibitem{xi} Note that the upper critical field extracted from resistivity is substantially larger than what is deduced here from the shift in the bulk transition temperature with magnetic field. This means that if one takes the resistive transition as reference\cite{Xiao}, the superconducting coherence length would be underestimated by a factor of two.
\end{thebibliography}
\end{document}